\documentstyle[12pt,epsf]{article}
\newcommand{\be}{\begin{equation}}
\newcommand{\ee}{\end{equation}}
\newcommand{\bear}{\begin{eqnarray}}
\newcommand{\ear}{\end{eqnarray}}

\newsavebox{\LSIM}
\sbox{\LSIM}{\raisebox{-1ex}{$\ \stackrel{\textstyle<}{\sim}\ $}}
\newcommand{\lsim}{\usebox{\LSIM}}
\newsavebox{\GSIM}
\sbox{\GSIM}{\raisebox{-1ex}{$\ \stackrel{\textstyle>}{\sim}\ $}}

\begin{document}
\begin{titlepage}
\begin{flushright}
HD-THEP/99-19\\
MZ-TH/99-22
\end{flushright}
\vspace{1cm}
\begin{center}
\vspace{1cm}
{\bf\LARGE A Tachyonic Gluon Mass:}\\
\vspace{.3cm}
{\bf \LARGE between Infrared and Ultraviolet}\\
\vspace{2cm}
Stephan J. Huber$^a$\\
Martin Reuter$^b$\\
Michael G. Schmidt$^a$\\
\vspace{.5cm}
$^a$ Institut f\"ur Theoretische Physik der
Universit\"at Heidelberg,\\
Philosophenweg 16, D-69120 Heidelberg, Germany\\
$^b$ Institut f\"ur  Physik,  Universit\"at Mainz,\\
 Staudingerweg 7, D-55099 Mainz, Germany
\\
\end{center}
\bigskip\noindent
\vspace{1cm}
\begin{abstract}
The gluon spin coupling to a Gaussian correlated background
gauge field 
induces an effective tachyonic gluon mass. It is 
momentum dependent and vanishes in the UV only like $1/p^2$. 
In the IR, we obtain stabilization through a positive 
$m^2_{\rm conf}(p^2)$ related to confinement. Recently a
purely phenomenological tachyonic gluon mass was used  to
explain the linear rise in the $q\bar q$ static potential at small 
distances and also some long standing discrepancies found 
in QCD sum rules.
We show that the stochastic vacuum model of QCD predicts a
gluon mass with the desired properties.
\end{abstract}
\end{titlepage}

Recently  it was argued \cite{1,2} that an effective negative gluon  
(mass)$^2$ produces the linear rise of the static quark-antiquark potential   
observed in lattice calculations \cite{3} at rather short distances below the  
nonperturbative IR scale of QCD. This can also be  related to an  
unconventional $Q^{-2}$ piece in the running QCD coupling $\alpha_s(Q^2)$  
\cite{6,7}
indicated again by lattice data on the gluon condensate and the 3-gluon  
vertex \cite{8}.  Later in ref. \cite{5}, it was shown that a linear term in  
the potential  for small distances appears in the evaluation of the Wilson  
loop integral in  a Gaussian correlated background gauge field if the  
paramagnetic (spin related) gluon coupling is taken into account. A gluon  
propagator with a nonperturbative  tachyonic mass in the evaluation of the  
operator product expansion for QCD sum rules also solves old discrepancies in  
the pion and scalar gluonium channels \cite{2}.

In ref. \cite{2}, the view was taken that a gluon mass is not just a  
convenient way to parametrize nonperturbative infrared (IR) effects, but that a  
tachyonic gluon mass also enters into the basic short distance operator product  
expansions. In this letter, we propose  less radically that a tachyonic gluon  
mass does appear  at momenta above the QCD scale, although at very high  
momenta/small distances it goes to zero. Thus, in the real ultraviolet regime (UV), the  
basic perturbative operator product expansion should be  
valid.\footnote{Concerning a $1/Q^2$ behavior consider the following simple  
function: $(\alpha Q^4+\beta Q^2)^{-1}$. If $\beta\gg\alpha$ is large, it  
behaves like $\frac{1}{\beta Q^2}$ for $Q^2<\beta/\alpha$, i.e. for rather  
large $Q^2$!} Of course, if one discusses an operator product expansion in an  
effective theory including a tachyonic gluon mass in which very high momenta  are  
already integrated out, this mass shows up in the expansion. However, at very  
small distances, the expansion loses its basis.

In the IR, a tachyonic gluon mass would cause divergences. Thus there should  
be an overcompensating positive effective gluon (mass)$^2$. In the case of 
the hot electroweak theory close to the  symmetric phase, we have argued  
earlier \cite{9} for the following picture: a tachyonic gauge boson mass  
related to spin-spin forces is effective at intermediate momenta and a  
positive (mass)$^2$ related to the area law of confinement is dominant in the  
gauge boson propagator in the IR. Our equations in ref. \cite{9} were written  
for a general dimension $d$ and gauge group $SU(N)$, but we had restricted our  
discussion to the case $d=3$ and $N=2$, which is the relevant one for the
discussion of the electroweak phase transition. 
In this letter, we analyze the QCD case $d=4$ 
and $N=3$.

Our actual calculations use the ``stochastic vacuum model'' \cite{10} for the  
QCD vacuum. In brief, this model provides the area law of confinement  
in a very simple and intuitive way.
It describes the QCD vacuum by a 
stochastic process and assumes that only Gaussian correlations are 
present in the cumulant expansion. 
The nonperturbative  part of the gauge 
invariant field strength correlator is given by (after contracting the Lorentz indices):
\be\label{1}
\langle\langle \hspace{.05cm} g^2 F_{\mu\nu}^a(x',x_0)F^b_{\mu\nu}(x,x_0)\rangle\rangle 
=\delta^{ab}\langle g^2 F^2\rangle D\left(\frac{(x'-x)^2}{a^2}\right).
\ee
Here $\langle g^2F^2\rangle\equiv\langle g^2F^a_{\mu\nu}F^a_{\mu\nu}\rangle$
is the usual ``local'' gluon condensate, and $a$ denotes the  correlation
length. The tensor $F_{\mu\nu}(x,x_0)$ is the ordinary field strength tensor
$F_{\mu\nu}(x)$ parallel-transported to a fixed common reference point $x_0$
\cite{10} and a Euclidean metric is used. A typical form of $D$ used in lattice
evaluations \cite{gia} is $\exp(-|x'-x|/a)$. 

The inverse gluon propagator (in the Feynman background gauge) in a 
background gauge field reads
\be\label{2}
K_{\mu\nu}^{ab}=\left[ -{l}\!\!D^2  \delta_{\mu\nu}+m^2 1\!\!{\rm l}_c\delta_{\mu\nu}
 +2ig{l}\!\!F_{\mu\nu}
\right]^{ab}.\ee
Here we have allowed for the presence of a gluon mass for future convenience. Linearized
gauge field excitations $a_{\mu}^b(x)$ interact with the background field via 
the Lagrangian $\frac{1}{2}a_{\mu}^bK_{\mu\nu}^{bc}a_{\nu}^c$. We
have to distinguish the ``diamagnetic'' coupling via the ordinary 
minimal-substitution term ${l}\!\!D^2\delta_{\mu\nu}$ and a non-minimal 
``paramagnetic'' coupling via the $2igF_{\mu\nu}$ term. In \cite{9} we used this 
interaction Lagrangian in order to compute the ``mass operator'' $\Sigma$
of the gluon (vacuum polarization) to lowest nontrivial order in 
$\langle g^2F^2 \rangle$. It is defined such that 
\be\label{prop}
G=[G^{(0)-1}-\Sigma]^{-1}
\ee 
where $G^{(0)}$ and $G$ are the free and the dressed propagator of the gluon, respectively. 
The latter describes the  propagation of a perturbative gluon in the 
background of a Gaussian-correlated random Yang-Mills field. Taking 
$x_0=(x+x')/2$ as the reference point, the result for $\Sigma$ is diagonal
in the Lorentz and the color indices:
\be
\Sigma(x,x')^{ab}_{\mu\nu}={\cal S}(z^2\equiv(x-x')^2)\delta_{\mu\nu}\delta^{ab}.
\ee 
The function ${\cal S}(z^2)\equiv{\cal S}_F(z^2)+{\cal S}_A(z^2)$ receives 
contributions from two rather different physical mechanisms: ${\cal S}_F$
results from the spin-related paramagnetic interactions via the $F_{\mu\nu}$
term alone, while ${\cal S}_A$ stems from diagrams involving the diamagnetic 
coupling and coincides with the result that we would obtain for spin-0 bosons. 
Fourier-transforming with respect to $z$, we obtain 
$\tilde{\cal S}(p^2,m^2)\equiv\tilde{\cal S}_F+\tilde{\cal S}_A$ with\footnote{See eq.
(A.34) of \cite{9}.}
\bear\label{3}
\tilde {\cal S}_F(p^2,m^2)&=&\frac{3}{32\pi^3}
\langle g^2 F^2\rangle\int^\infty_0 dq
q^3\tilde D(q^2)\int^{+1}_{-1}\frac{d\cos\vartheta(1-\cos^2\vartheta)^{1/2}}
{(p^2+q^2+m^2-2pq\cos\vartheta)}\nonumber\\
&=&\frac{3}{128\pi^2}\langle g^2 F^2\rangle\int^\infty_0 dq
q^3\tilde D(q^2)\times \nonumber\\
&&\frac{1}{(pq)^2}\left[(p^2+q^2+m^2)-\sqrt{(p^2+q^2+m^2)^2-(2pq)^2}
\hspace{.1cm}\right]
\ear
and similarly for $\tilde{\cal S}_A(p^2,m^2)$.

In view of eq. (\ref{prop}) it is natural to interpret $m^2_{\rm tach}(p^2)\equiv
-\tilde{\cal S}(p^2,m^2=0)$ as the effective, momentum-dependent 
mass of a gluon propagating in a stochastic gauge field background. 
One finds that $\tilde{\cal S}(p^2,0)>0$ so that $m^2_{\rm tach}<0$, 
i.~e.~the effective mass is {\it tachyonic}. Its value at $p^2=0$,
$m^2_{\rm tach}\equiv m^2_{\rm tach}(0)$, can be written down in
closed form\footnote{Cf.~eq. (A.38) of \cite{9} where $w\equiv z^2/a^2$.}:
\be\label{4}
-m^2_{\rm tach}=\tilde {\cal S}(p^2=0,m^2=0)=\frac{3}{32}a^2\langle g^2 F^2\rangle
\{1+\delta\}\int^\infty_0 dw\ D(w)
\ee
Here the ``1'' inside the curly brackets on the RHS of eq. (\ref{4}) comes from
the paramagnetic ${\cal S}_F$ while
\[\delta=\frac{17}{12}-\frac{4}{3}\ln 2\approx -0.492 \]
is the contribution of the diamagnetic piece ${\cal S}_A$. We observe that the
tachyonic nature of the induced mass is a consequence of the paramagnetic
spin interaction which dominates over the diamagnetic one. The latter generates
a (mass)$^2$ which is numerically smaller and of opposite sign.

It is quite remarkable that, up to a constant factor, the RHS of eq. (\ref{4})
equals precisely the well-known \cite{10} result of the stochastic
vacuum model for the string tension (with the color sources in the fundamental
representation of $SU(3)$): 
\be\label{5}
\sigma_{\rm fund}=\frac{\pi^2}{144}a^2\langle g^2F^2\rangle\int^\infty_0 dwD(w).
\ee
Thus the gluon mass can be expressed directly in terms of the string tension:

\be\label{6}
-m^2_{\rm tach}=\frac{18}{\pi}(\ln 2-\frac{5}{16})\sigma_{\rm fund}
\approx 2.18\sigma_{\rm fund}.
\ee
It is a universal prediction of the stochastic vacuum model in the 
sense that no separate knowledge of the gluon condensate, the 
correlation length
and the shape of the structure function $D$ is needed.
In fact, in eq. (\ref{6}) we can directly use the experimental
value for the string tension. From the charmonium spectrum
one obtains \cite{tens} $\sigma_{\rm fund}\approx (430 MeV)^2$
which leads to $m^2_{\rm tach}\approx -0.40(GeV)^2$ or 
$|m_{\rm tach}|\approx 635MeV$. 

This is indeed a tachyonic mass of the order of magnitude 
postulated in ref. \cite{2} on phenomenological grounds.
Since to lowest order 
\be \label{5b}
G_{\mu\nu}^{ab}(p^2)=\left[\frac{1}{p^2}-\frac{m^2_{\rm tach}}{p^4}
+{\cal O}\left(\frac{1}{p^6}\right)\right]\delta_{\mu\nu}\delta^{ab},
\ee
with an $1/p^4$ correction to the free propagator, we see that in the
static $q\bar q$ potential $V(r)$ the tachyonic gluon mass gives  rise to a 
term that grows linearly  with $r$. However, in contrast to the asymptotic
($r\rightarrow\infty$) regime, in which $V(r)\approx\sigma_{\rm fund }r$
dominates the Coulomb term, we are at much shorter
distances here where the linear term is only a small correction to the 
$1/r$ law.

\begin{figure}[t]
\begin{picture}(200,150)
\put(-150,-20){\epsfxsize7cm \epsffile{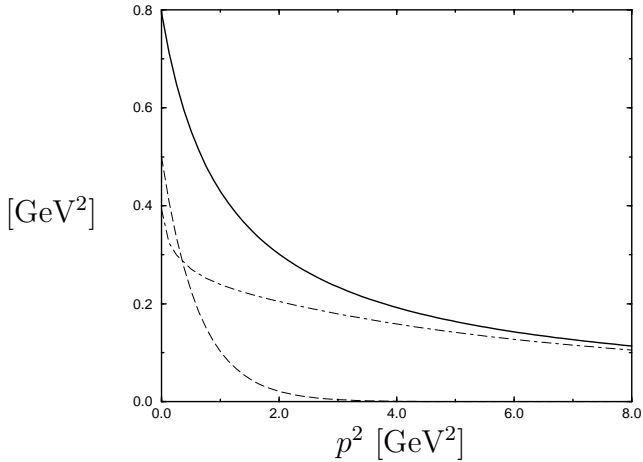}}
\put(175,-7){$p^2$ [GeV$^2$]}
\put(50,80){[GeV$^2$]}
\end{picture} 
\caption{${\cal S}_F(m^2_{\rm conf}=0)$ (solid line), 
${\cal S}_F(m^2_{\rm conf}\neq 0)$ (dashed-dotted line) and
$m^2_{\rm conf}$ (dashed line) as function of $p^2$.}
\label{fig1} 
\end{figure}

In fig. 1, we present $\tilde {\cal S}_F(p^2,m^2=0)$ for  
$D(w)=\exp(-|z|/a)=\exp(-|w|^{\frac{1}{2}})$ and $a=0.22$fm \cite{gia}
as a function of $p^2$. It vanishes for large $p^2$ rather slowly as
\be\label{7}
\tilde {\cal S}_F(p^2)\begin{array}{c}{}\\ \longrightarrow\\
p^2\gg\frac{1}{a^2}\end{array}\frac{3}{8}\frac{\langle g^2F^2\rangle}{p^2}.\ee

Eqs.~(\ref{3}), (\ref{6}) and (\ref{7}) are our main results. 
The ``paramagnetic dominance'' and the 
resulting tachyonic sign of the 
gluon (mass)$^2$ are related to the well known instability of the 
Savvidy vacuum \cite{sav}. The latter models the true QCD vacuum in
terms of an external gauge field with a constant field strength. While
already lower in energy than the perturbative vacuum, it is unstable
and decays towards the true ground state precisely because the destabilizing
paramagnetic interaction of the vacuum fluctuation overrides the diamagnetic
one. The paramagnetic dominance encountered above (and discussed in detail
in ref. \cite{9}) is a remnant of this phenomenon for non-constant 
backgrounds. 

The self-energy $\Sigma$ can also be used in order to understand why,
first of all, the perturbative vacuum is unstable. Inserting $\Sigma$, which
is proportional to $\langle g^2F^2\rangle$, into a perturbative gluon loop,
we obtain the $\langle F^2\rangle$ term of the 1-loop effective potential
for the condensate, $V_{\rm eff}(\langle F^2\rangle)$. This term turns out to be
negative, i.~e.~$\langle g^2F^2\rangle=0$ is not the minimum of $V_{\rm eff}$,
and a non-zero condensate tends to form.   
But, in contrast to the $d=3$ case, for $d=4$, the  term that is linear in 
$\langle F^2\rangle$ is logarithmically divergent in the UV.

In ref. \cite{9}, we also calculated the standard perturbative 1-loop  
gluon self-energy $\Pi$  and we introduced the $F_{\mu\nu}$ correlator (\ref{1}) 
only in a second step (fig. 2).
This should be an equivalent procedure and indeed, in three dimensions, this  
can be easily checked. For $d=4$, the perturbative vacuum polarization $\Pi(q^2)$  
(fig. 2) is divergent. 
It has to be renormalized at some Euclidean $q^2=\mu^2$  in the 
perturbative range $\mu^2>1/a^2$.\footnote{According to  ref. \cite{dnr} the
stochastic vacuum model describes the QCD vacuum at a renormalization
scale $\mu$  not much  above $1/a$.} Taking only the contribution from
the non-minimal gauge boson coupling, one obtains
(see eq. (2.20) of ref. \cite{9})
\be 
\Pi_F^{\rm ren}(p^2,\mu^2)=\frac{4}{(4\pi)^2}\log\frac{p^2}{\mu^2} 
\ee
which is to be inserted into
\be 
V_{\rm eff}=\frac{3}{4}\langle g^2 F^2\rangle\int\frac{d^4p}{(2\pi)^4}
\Pi_F^{\rm ren}(p^2,\mu^2)\tilde D_{\rm eff}(p^2).  
\ee
This is indeed negative if $\mu^2>1/a^2$, the typical scale set by $\tilde D$.

\begin{figure}[t]
\begin{picture}(200,140)
\put(520,100){\epsfxsize5cm \epsffile{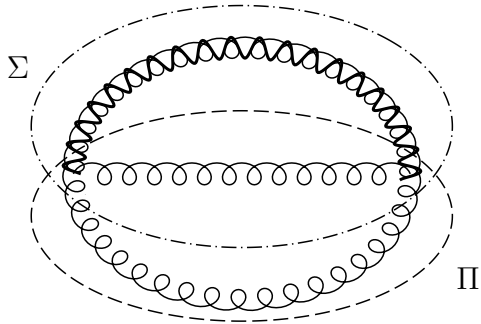}}
\put(270,20){$\Pi$}
\put(100,100){$\Sigma$}
\end{picture} 
\caption{First order contribution of the correlator (double wiggled)
to the gluon loop}
\label{fig2} 
\end{figure}

The 1-loop expression resulting from the $\Sigma$ approach,
\be\label{7a}
V_{\rm eff}\propto\int\frac{d^4p}{(2\pi)^4}\{\log(p^2-\tilde {\cal S}_F(p^2))-\log p^2\},\ee
with an appropriate constant already subtracted, is still  
logarithmically divergent.  
Obviously the $1/p^2$ decay of $\tilde{\cal S}(p^2)$ is too slow to make
$\Sigma$ completely irrelevant in the UV. At first sight, this seems to
be an UV-effect related to the gluon mass similar to the one discussed
in ref. \cite{2}. However, our divergent term is proportional to the 
stochastic average of the classical action $\frac{1}{4}F_{\mu\nu}^aF_{\mu\nu}^a$
and can be removed by the same renormalization as in
perturbation theory. This is an important difference as compared to ref. \cite{2}.
Note also that in contrast to the case with a constant background  field,
we are not plagued by IR singularities any longer.

To really obtain an instability at $\langle F^2\rangle=0$, the positive 
tree term has to be dominated by the quantum corrections. 
The 1-loop potential could be further improved by including  
Wilson type renormalization effects \cite{11}: the effective $g^2(k^2)$ is 
increasing  in the infrared direction $k^2\to 0$.

Finally, let us discuss the gluon propagator and the stochastically averaged
1-loop effective potential 
\be \label{10a}
V_{\rm eff}(\langle F^2\rangle)=\frac{1}{2} \langle\langle\mbox{Tr log}(K)
\rangle\rangle
\ee
beyond the linear approximation in $\langle F^2\rangle$. The effective potential
can be represented in terms of a world-line path integral \cite{9} over
closed paths $y(\tau)$. Its integrand contains the pivotal factor
\[\langle\langle{\rm tr}_{\rm cL}P\exp{ig\int d\tau(\dot y_{\mu} A_{\mu}(y)+2F(y))}
\rangle\rangle.\]
Apart from the paramagnetic $F$-term discussed above, this is
precisely the stochastic average of the Wilson loop operator
which signals confinement. For large loops, it behaves as $\exp(-\sigma
{\cal A}[y]T)$ where $\sigma\equiv\sigma_{\rm fund}\propto \langle g^2F^2\rangle$
is the string tension (\ref{5}) and ${\cal A}[y]$ denotes the area of the
minimal surface bordered by $y(\tau)$. This is one of the most
important results of the stochastic vacuum model. Thus the
contribution to $V_{\rm eff}$ which originates from the ``diamagnetic''
interaction of very large loops is proportional to
\be\label{8}
V_{\rm eff}^{\rm conf}=-\frac{1}{2}\int^\infty_0\frac{dT}{T} T^{-2}\int Dy
\exp(-\int^1_0 d\tau \dot y^2/4)
\{\exp\left(-\sigma {\cal A}[y] T f({\cal A}[y]T)\right)-1\}.
\ee
Here we use appropriately rescaled dimensionless variables. The path integration is
subject to the conditions $y(0)=y(1)$ and $\int_0^1y(\tau)d\tau=0$. In principle,
the cut-off function $f$ with $f(\infty)=1$ and $f(0)=0$ could be calculated from the
model; it serves the purpose of cutting out the contribution of small loops so that
the perturbative expression at small distances is not changed. Using the identity
$(4\pi T)^{-2}=\int\frac{d^4p}{(2\pi)^4}\exp(-Tp^2)$, the confinement-related 
part of $V_{\rm eff}$ assumes the form\footnote{One of the authors (M.~G.~Schmidt)
thanks M.~Laine for a helpful discussion on that point.} 
\be \label{11a}
V_{\rm eff}^{\rm conf}(\langle F^2\rangle)=\int\frac{d^4p}{(2\pi)^4}I(p^2)
\ee
where 
\be\label{9}
I(p^2)=-4\pi\int^{+\infty}_{-\infty}d\omega
F(\omega)\int^{\infty}_0\frac{dT}{T^2}
\exp(-p^2 T) G\left(\frac{\omega}{T}\right)\ee
with
\be\label{10}
F(\omega)=\int Dy\exp\left\{ i {\cal A}[y]\omega-\int^1_0 d\tau \dot  
y^2/4\right\}\ee
and
\be\label{11}
G(x)=2\int^\infty_0 d\bar A\cos (\bar A x)\left[\exp\{-\sigma
\bar A f(\bar A)\}-1\right].\ee

Expression (\ref{9}) can be written in the form $\{\ln(p^2+m^2_{\rm conf}(p^2))-\log
p^2\}$. Eq. (\ref{11a}) suggests that 
this defines an effective IR mass $m^2_{\rm conf}(p^2)$ due to  
confinement \cite{9}. Note that the counterterm cancels in this relation. 

The  evaluation of
(\ref{9}), (\ref{10}) and (\ref{11}) requires numerical studies, in particular of 
the most interesting function $F(\omega)$ \cite{12}. The main scale in this  
problem is the string tension $\sigma$; thus we expect $m^2_{\rm conf}(p^2=0)\sim  
\sigma$. The shape of the cut-off function $f(\bar A)$ determines the profile  
in $p^2$. At $p^2=0$, $m^2_{\rm conf}$ should dominate the negative $-\tilde  
{\cal S}_F(p^2=0)$ in order to  obtain a reasonable IR behavior (though interferences between  
confinement and spin effects  are not treated carefully in the approach  
presented above).

The confinement mass $m^2_{\rm conf}(p^2)$ should be also included in our 
calculation of $\tilde  
{\cal S}_F$. In a first approximation, one could substitute  it as an $m^2$ in formula (\ref{3})  
(of course, it is again not an ultraviolet mass requiring regularization). This  
lowers $\tilde {\cal S}_F$ and the result depends a lot on the balance between  
$m^2_{\rm conf}$ and $-\tilde {\cal S}_F$ at small $p^2$. In fig. 1, we have included a  
plausible function $m^2_{\rm conf}(p^2)$  to demonstrate the effect. 
We expect $m^2_{\rm conf}(p^2)$ to dominant in the IR, i.~e.~for $p^2\lsim1/a^2$.  
The range of $\tilde{\cal S}_F(p^2)$ extends to higher $p^2$.  
It vanishes only like $1/p^2$. Thus $-\tilde{\cal S}_F(p^2)$ mimics a  
tachyonic UV mass.

\section*{Acknowledgment}
We would like to thank H.~G.~Dosch, A.~Hebecker, S.~Klevansky and S.~Narison 
for useful discussions
and Yu.~Simonov for referring us to  \cite{2} and \cite{5}. This work was 
supported in part by the TMR network {\it Finite Temperature Phase 
Transitions in Particle Physics}, EU contract no. ERBFMRXCT97-0122.

\end{document}